\documentstyle[preprint,aps,epsf,prd,eqsecnum]{revtex}

\def\pubnumber{\vbox{\hbox{KUNS
1461}}}

 \def\scri{\hbox{${\cal J}$\kern -.645em {\raise
      .57ex\hbox{$\scriptscriptstyle (\ $}}}}

\newcommand{\remark}{\noindent{\bf Remark:}\ }

\newtheorem{Theorem}   {Theorem}   [section]
\newtheorem{Corollary} [Theorem]   {Corollary}
\newtheorem{Lemma}     [Theorem]   {Lemma}
\newtheorem{Proposition} [Theorem] {Proposition}

\begin{document}
\parbox{\hsize}{ 
\rightline{\pubnumber}
\title{Stable Topologies of Event Horizon}
\author{Masaru Siino
\\
\it Department of Physics, Kyoto University\\
Kitashirakawa, Sakyoku, Kyoto 606-01, Japan}
\maketitle
}
\begin{abstract}
In our previous work, it was shown that the topology of an event horizon 
(EH) is determined by the past endpoints of the EH. A torus EH (the 
collision of two EH) is caused by the two-dimensional (one-dimensional) 
set of the endpoints. In the present article, we examine the stability of 
the topology of the EH.
We see that a simple case of a single spherical EH is unstable. 
Furthermore, in general,
an EH with handles (a torus, a double torus, ...) is structurally stable in the 
sense of catastrophe theory.
\end{abstract}

\leftline{PACS number(s): 02.40.-k, 04.20.Gz, 05.45}
\section{Introduction}
The topology of an event horizon (EH) is very important when one 
investigate the various properties of the EH, and is sometimes
considered to be 
trivial. For example, one may assume that the 
topology of the event horizon (TOEH) is a sphere for the 
uniqueness theorem of a black hole. On the other 
hand, it is natural that the TOEH is a sphere in an 
astrophysical sense. Furthermore, 
many authors\cite{HW} proved that the TOEH is a sphere under some conditions.

On the contrary, the present author have shown the TOEH is determined by the structure of
the endpoints of the EH\cite{MS}. From this, the two-dimensional 
(one-dimensional) set of the endpoints is related to an EH  with a 
torus topology (the collision of the EH).
Therefore the question what determines the structure of the endpoints, 
arises. To discuss this problem, we need to study the dynamics 
of this structure. For this reason, it is worth determining
 the stability of the structure of the endpoints. 

Hence, the purpose of the present article is to investigate the stability of
the structure of the endpoints. From this investigation, we will find the 
stability of the TOEH.

First we investigate the stability of a spherical EH under linear 
perturbation. 
Especially, we examine the causal structure of 
a perturbed Oppenheimer-Snyder spacetime to discuss the stability of its 
endpoint. Second,
catastrophe theory is applied to the EH for more general discussion.
 We argue the structural 
stability of 
more general cases than the spherical EH.

In the next section, we briefly introduce the our previous work\cite{MS}, 
while the proof of the result are not given in this article. The 
third section shows the discussion of linear perturbation in a 
spherically symmetric spacetime. In the section 4, the structural 
stability of the endpoints is investigated with the base of catastrophe 
theory. The final section provides summary and discussions.
\section{The topology of event horizons}
In this section, we briefly introduce only the result of our previous 
work\cite{MS}.
Now we apply the theories of topology change\cite{SR}\cite{GR} to 
EHs.
Let $(M,g)$ be a four-dimensional $C^\infty$ spacetime whose topology is 
$R^4$.
In the rest of this article, the spacetime  $(M,g)$ is supposed to 
be strongly causal. Furthermore, 
for simplicity 
the topology of the EH
(TOEH\footnote{The TOEH means the topology of the spatial section of the EH. 
Of course, it
depends on a timeslicing.}) is assumed to be a smooth $S^2$
far in the future and the EH is not eternal one (in other words, the EH 
begins somewhere in the spacetime, and is open to the infinity in the 
future direction with a smooth $S^2$ section).
These assumptions will be valid when we consider only one regular 
($\sim R\times S^2$)
asymptotic region, namely the future null infinity 
$\scri^+$, to define the EH,
and the formation of a black hole. 

In our investigation, the most important concept is the existence of 
the endpoints of
 null geodesics $\lambda$ which completely lie in the EH and generate it. We call them 
 the endpoints of the EH.
To generate the EH the null geodesics $\lambda$ are maximally extended to 
the future and past as long as they belong to the EH. Then the endpoint 
is the point where such null geodesics are about to come into the EH (or go 
out from the EH), though the null 
geodesic  can continue to the outside or the inside of the EH through 
the    endpoint in the sense of the whole spacetime.
 We consider a null vector field $K$ on the EH which is tangent to the null 
 geodesics $\lambda$.
$K$ is not affinely parametrized but parametrized so as to be continuous 
even on the endpoint where the caustic of $\lambda$ appears. Then the endpoints of $\lambda$ are the zeros of 
$K$, which can become only past endpoints since $\lambda$ must reach to 
infinity in the future direction. 

First we pay attention to the relation between 
the endpoint and the differentiability of the EH. We see that
the EH is not differentiable at the past endpoint.    

\begin{Lemma}\label{L1}
Suppose that $H$ is a three-dimensional null surface imbedded into the spacetime $(M,g)$ by 
a function $F$ as
\begin{equation}
H:x^4=F(x^i,i=1,2,3),
\end{equation}
in a coordinate neighborhood 
$({\cal U}_{\alpha},\phi_\alpha),\ \phi_\alpha:{\cal U}_{\alpha}\rightarrow 
R^4$, where 
$\partial/\partial x^4$ is timelike.
When $H$ is generated by the set of null geodesics  whose tangent vector 
field is $K$,
$H$ and the imbedding function $F$ are indifferentiable at the endpoint of the null
 geodesic (the zero of $K$).
\end{Lemma}

Here, 
we assume that the EH is $C^r(r\ge 1)$-differentiable except on the 
endpoint of the null geodesics generating the EH and the set of the 
endpoints is compact. Thus we suppose that the EH is indifferentiable 
only on a compact subset. 

Next, we prepare a basic proposition.
Suppose there is no past endpoint of the null geodesic generator of an EH between 
$\Sigma_1$ and $\Sigma_2$. 
 Then, Geroch's theorem\cite{MS}\cite{GR} stresses  the topology of the smooth
EH does not change. 

\begin{Proposition}  \label{P1}
Let $H$ be the compact subset of the EH of $(M,g)$, whose boundaries are an initial spatial section
$\Sigma_1$ and a final spatial section $\Sigma_2$, 
$\Sigma_1\cap\Sigma_2=\emptyset$. $\Sigma_2$ is assumed to be
far in the future and a smooth sphere. Suppose that $H$ is $C^r(r\ge 
1)$-differentiable. Then
the topology of $\Sigma_1$ is $S^2$.
 \end{Proposition}

Now we discuss the possibilities of non-spherical topologies.
From Sorkin's theorem there should be any zero of null vector field
 $K$ in the interior of $H$ provided that
the Euler number of $\Sigma_1
$ is different from that of $\Sigma_2\sim S^2$. Such a zero  can only be the past endpoint of the EH since the null 
geodesic generator of the EH
cannot have a future endpoint. About this past
 endpoint of the EH we state 
the following proposition.

\begin{Proposition}\label{P2}  
           The set of the past endpoints (SOEP) 
 of the EH 
is a connected spacelike set.
\end{Proposition}

Then, we give theorems and corollaries about the topology of the 
spatial section of the EH on a timeslicing.
First we consider the case where the EH has simple structure.

\begin{Theorem}\label{T1}
Let $S_H$ be the section of an EH 
by a spacelike hypersurface.
If the EH is $C^r(r\ge 1)$-differentiable at $S_H$, it is
topologically $\emptyset$ or $S^2$.
\end{Theorem}

On the other hand, we get the following theorem about the change of the  
 TOEH with the aid of Sorkin's
 theorem\cite{SR}. 
\begin{Theorem}\label{T2}
Consider a smooth timeslicing ${\cal T}={\cal T}(T)$ defined by a smooth 
function $T(p)$;
\begin{equation}
        {\cal T}(T)=\{p\in M\vert T(p)=T=const.,\ \  T\in 
        \left[T_1,T_2\right]\}, \ \ g(\partial_T,\partial_T)<0.
        \label{}
\end{equation}
Let $H$ be the subset of the EH cut by ${\cal T}(T_1)$ and ${\cal T}(T_2)$, 
whose boundaries are the initial spatial section $\Sigma_1\subset 
{\cal T}(T_1)$ and the final spatial section $\Sigma_2\subset 
{\cal T}(T_2)$, and $K$ be the null vector field generating the EH.
 Suppose that $\Sigma_2$ is a 
sphere.  If, 
in the timeslicing $\cal T$, the TOEH changes ($\Sigma_1$ is not
homeomorphic to $\Sigma_2$) then there is the SOEP (the zeros of $K$) in 
$H$, and when the timeslice touches
\begin{itemize}
        \item the one-dimensional segment of the SOEP, it causes the coalescence of two
        spherical EHs.
        \item  the two-dimensional segment of the SOEP, it causes the change of the TOEH
        from a torus
        to a sphere.
\end{itemize}        
\end{Theorem}

This theorem needs the following remark.

\remark
One may face special situations.
 The possibilities of the branching 
endpoints should be noticed. If the SOEP
possesses a branching point, a special timeslicing can make the branching 
point into a point where the TOEH changes though such a timeslicing loses this aspect under 
the small deformation of the timeslicing. The index
of this branching endpoint may deny a direct consideration.
 The situation, however, is regarded as the degeneration of the two 
 distinguished SOEP.
 Imagine a little slanted timeslicing, and it will decompose the branching point
 into two distinguished (of course, there are the possibilities of the degeneration of three or more)
SOEP.  Some of examples are shown in the following. The first case is
 the branch of
  the one-dimensional SOEP\footnote{We can also
   treat the branching   points of the two-dimensional SOEP in the same
   manner.}. Then, three
   spheres coalesce there. The next case is a one-dimensional branch from
    the two-dimensional SOEP.
This branching point is the degeneration of 
 the one-dimensional
SOEP and the two-dimensional SOEP.
This decomposition tells that the TOEH changes at
 this point, for example, from a sphere and a torus to a sphere.

\begin{figure}
        \centerline{\epsfxsize=16cm \epsfbox{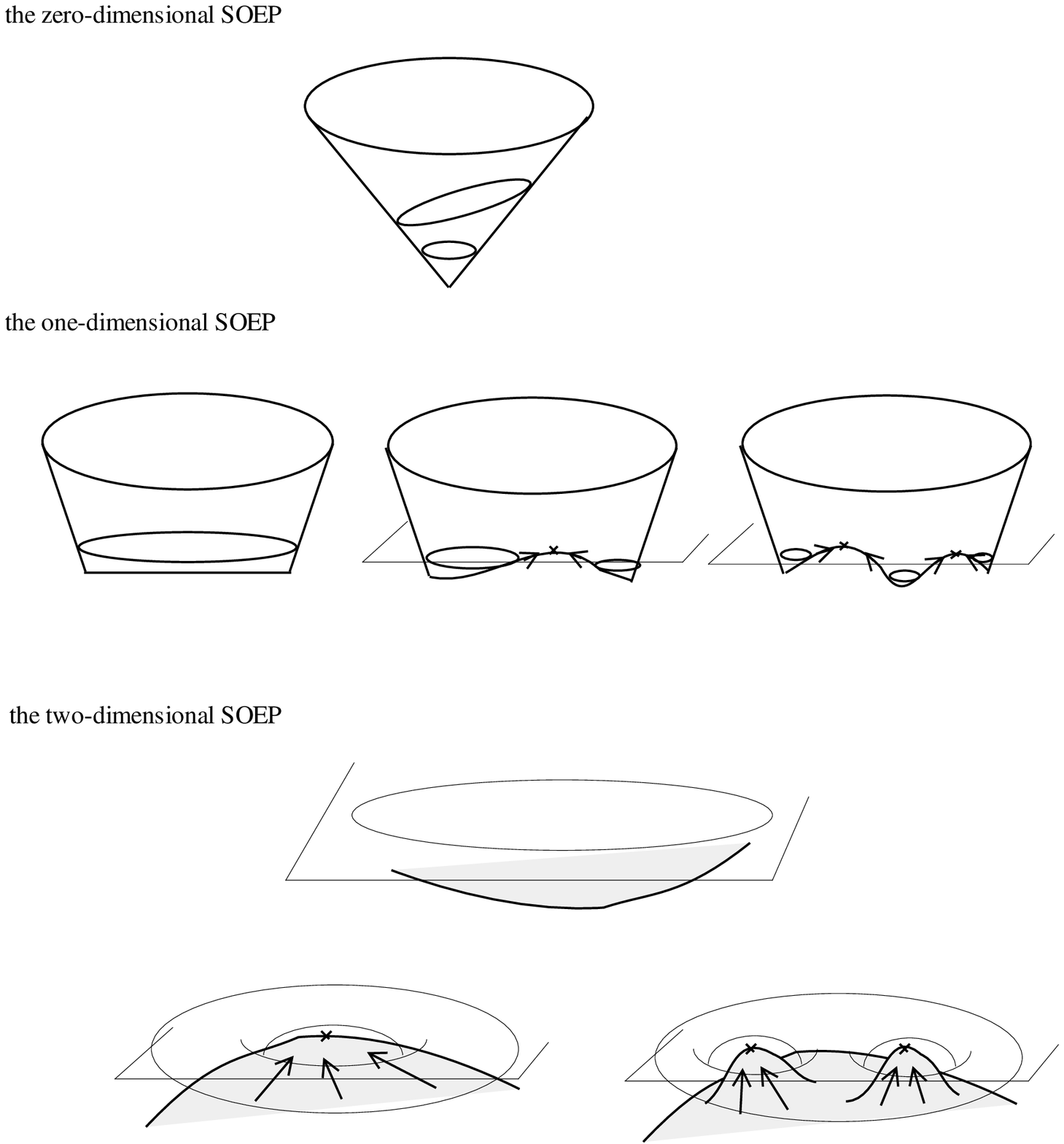}}
        \caption{EHs with the zero-, one- and two-dimensional SOEP are 
        shown. We see that the one-dimansional SOEP becomes coalescence 
        of arbitrary number of spherical EHs. For the two-dimensional 
        SOEP, only sections of the EH and the SOEP are drawn. It can 
        become the EH with arbitrary number of handles. It is also 
        possible to change the EH into the trivial criation of a spherical EH.}
        \protect\label{fig:ndend}
\end{figure}

Incidentally,
 a certain timeslicing gives the further changes of the Euler number (see 
 Fig.\ref{fig:ndend}). 
\begin{Corollary}\label{CL1}
The topology changing processes of an EH from $n\times S^2$
to $S^2$ ($n=1,2,3,...$) can change each other,   and 
from a surface with genus=$n$  to $S^2$ ($n=1,2,3,...$) can also change,
  under the  appropriate
deformation of their timeslicing. 
\end{Corollary}
   
As shown in the corollary \ref{CL1}, the TOEH  highly depends on the 
timeslicing. Nevertheless, the theorem \ref{T2} tells that there is the 
distinct difference
between the coalescence of $n$ spheres  where the Euler number decreases by the
one-dimensional SOEP and the EH of a surface with genus=$n$ where the Euler number
increases by the two-dimensional SOEP.

Finally we see that, in a sense, the TOEH is a transient term (see 
 Fig.\ref{fig:ndend}).
\begin{Corollary}
          All the changes of the TOEH are reduced to the trivial creation of 
          an EH which is topologically $S^2$.
\end{Corollary}

Thus we see that the change of the TOEH is determined by the topology of 
the SOEP and the timeslicing way of it. To fix the TOEH we must only give the order to each
 vertex, edge or face of  the SOEP by a timeslicing.


\section{The linear perturbation of an event horizon with a spherical topology}
The purpose of this section is to investigate the stability of the TOEH 
which always is a spherical topology, under linear perturbation. From 
our work in the previous section\cite{MS},
such an EH has only one zero-dimensional SOEP (see Fig. \ref{fig:ndend}). Then, we investigate whether this 
zero-dimensional SOEP is stable under linear perturbation. Now, it
should be noted the `stable' does not mean that the perturbation does not blow
 up but
the TOEH is not changed by the perturbation.

 As a 
background spacetime, a spherically symmetric spacetime is appropriate. 
If the spherical symmetric spacetime has a non-eternal EH, it has only 
one endpoint at the origin and the TOEH is always a sphere. In this 
case, it is possible to study the linear perturbation in an established 
framework\cite{SG}.

Now we consider the Oppenheimer-Snyder spacetime as the familiar example 
of the EH with a spherical topology.
Its line element is given by

\begin{itemize}
        \item interior:
\begin{eqnarray}
        ds^2=a\left(\eta\right)^2\left(-d\eta^2+d\chi^2+\sin^2\chi 
        d\Omega^2\right),\ \ \ 0\le\chi\le\chi_0
        \label{eqn:int}\\
        a(\eta)=\frac12 a_m(1+\cos\eta),
\end{eqnarray} 
    \item       exterior:
\begin{eqnarray}
        ds^2 &=& -\left(1-2m/R\right)dt^2+{dR^2 \over 1-2m/R}+R^2 
        d\Omega^2,\ \ \ R_B(t)\le R \label{eqn:ext}\\
             &=& \left({32m^3\over R}\right)e^{-R/m}\left(-dV^2+dU^2\right)
             +R^2 d\Omega^2, 
\end{eqnarray}
\end{itemize}
where $V$ and $U$ are Kruskal-Szekeres coordinates.
When these geometries are continuated at $\chi=\chi_0$, the parameters of 
the exterior region are related to $a_m$ and $\chi_0$ as
\begin{eqnarray}
         m& = & \frac12 a_m\sin^3\chi_0,
        \label{¥} \\
         R_B& = & {a_m\sin\chi_0\over 2}(1+\cos\eta).
        \label{¥}
\end{eqnarray}

In the background spacetime the equations of null geodesics are easily 
solved and integrated. The background values of an outgoing null geodesic 
$\gamma$ in the direction $\theta_0, \phi_0$ and from the origin at $\eta=\eta_0$ are
\begin{eqnarray}
        l_0^a & = & \left({\partial \over \partial \eta}\right)^a+
        \left({\partial \over \partial \chi}\right)^a     \label{eqn:lin} \\
         & = & \left({\partial \over \partial V}\right)^a+
        \left({\partial \over \partial U}\right)^a       \label{eqn:lex} \\
        \gamma(\eta_0,\theta_0,\phi_0)&:&\ \ \ 
        \cases{\chi=\eta-\eta_0,\cr
        U-U_0(\chi=\chi_0,\eta=\chi_0+\eta_0)=
        V-V_0(\chi=\chi_0,\eta=\chi_0+\eta_0)\cr}\\
        \theta &=&\theta_0, \ \ \ \phi=\phi_0, \\
        \eta_{crit} & = & \pi-3\chi_0,
\end{eqnarray}
where $l_0$ is an outgoing null vector field and $\eta_{crit}$ is the 
supremum of the time $\eta$ when 
light ray emitted from the origin can 
reach to the future null infinity $\scri^+$. The SOEP of the EH is a point at 
the origin with $\eta=\eta_{crit}$. 

We expand the freedom of linear perturbation by spherical harmonics 
$Y_{LM}$, and they are decomposed into odd 
parity $[(-1)^{L+1}]$ modes and even parity $[(-1)^{L}]$ modes.  Since they are decoupled in the spherically 
symmetric background, we discuss the stability of the TOEH under each mode 
of the perturbation with a 
parity, $L$, and $M$. First we develop the property of null geodesics 
in a perturbed spacetime. The equations of null 
geodesics are given by,

\begin{eqnarray}
        0 & = & g_{ab} l^a l^b 
        \label{} \\
          & = & \left(g_{0ab}+h_{ab}\right)\left(l_0^a+\delta 
          l^a\right)\left(l_0^b+\delta l^b\right)
        \label{¥} \\
          & = & h_{ab} l_0^al_0^b + 2 g_{0ab} \delta l^a l_0^b,
        \label{¥}
\end{eqnarray}
and
\begin{eqnarray}
        l^a\nabla_a l^b & = & \alpha l^b
        \label{¥} \\
        l^a\partial_a l^b +  \Gamma^b_{ac} l^al^c&=&\alpha l^b \rightarrow 
        \label{¥} \\
        l_0^a\partial_a \delta l^b +\delta l^a\partial_a l_0^b+2\Gamma^b_{0ac} l_0^a \delta l^c+\delta\Gamma^b_{ac}
         l_0^a l_0^c& =&\delta\alpha l_0^b+\alpha_0\delta l^b,  
        \label{¥}      
\end{eqnarray}
where $g_0, \Gamma_0$ is given by (\ref{eqn:int}), (\ref{eqn:ext}) and 
$l_0$ (\ref{eqn:lin}), (\ref{eqn:lex}).
$\delta\alpha$ corresponds to the parametrization of $l$, and is set so 
that $\delta l^{\eta}+\delta l^{\chi}$ ($\delta l^V+\delta l^U$) vanishes. The deformation $\delta x^a$ of the light path $\gamma$ 
fixing its end on the same position of the future null infinity $\scri^+$,
  is integrated backward along the 
background light path $\gamma$ from the future null infinity to a point 
in the interior region as
\begin{eqnarray}
     \delta\alpha&=&\delta\Gamma^{\eta(V)}_{ab}l_0^al_0^b+\delta\Gamma^{\chi(U)}_{ab}l_0^al_0^b\\
         \delta \eta&=&-\delta \chi=\int^{\chi}_{\chi=\chi_0
         }d\chi[\gamma]{h_{\eta\eta}+2h_{\eta\chi}+h_{\chi\chi}\over 4 
         a^2}+\delta\eta(\chi=\chi_0),  \label{eqn:dec}\\
         \delta\eta(\chi=\chi_0)&=&{\partial\eta\over\partial V}\delta 
         V_0+{\partial\eta\over\partial U}\delta U_0, \\ \delta V_0=-\delta U_0&=&
         \int^{U_0}_{U=\infty 
         }dU[\gamma]{(h_{VV}+2h_{VU}+h_{UU})Re^{R/m}\over 32m^3},
        \label{}
\end{eqnarray}
since $\delta V_0=-\delta U_0$ implies $\delta\eta(\chi_0)=-\delta\chi(\chi_0)$.

\subsection{even parity mode}
The metric perturbation of the even parity mode is given by
\begin{equation}
        h_{ab}=
        \bordermatrix{
        &
         \eta,V\  \chi,U
        &
                        \theta\ \ \    \phi 
                        \cr
        &\bar{h}_{AB}Y_{LM}  &  \bar{h}_AY_{LM,\alpha}\cr
        &       Sym    &  r^2(K\gamma_{\alpha\beta}Y_{LM}+GY_{LM;\alpha\beta})\cr
        },
        \label{eqn:eve}
\end{equation}
where $r$ is a circumference radius and $\gamma_{\alpha\beta}$ is the 
metric of the 
unit sphere\cite{SG}.
For the even parity mode, the angular distribution of the $\delta\eta$ 
and  $\delta\chi$ (\ref{eqn:dec}) is 
just the spherical harmonics $Y_{LM}$. So, it is helpful to discuss the symmetry 
of each $Y_{LM}$.

 Since $Y_{00}$ is a spherically symmetric 
function it causes no change of the SOEP, unless
the perturbation is unstable and destroy the whole of the EH.
The even parity modes with $L=1$, $M =\pm 1$, can change into the mode of $Y_{10}$ 
by a certain rotation, and we only consider $M=0$ for $L=1$ mode. By
$Y_{10}$ perturbation, the wave front of light around 
the origin is shifted along the $z$-axis. Then, we only need to determine 
perturbed light paths starting from the origin for the north and the 
south. From eq. (\ref{eqn:eve}), we see 
$\delta\chi(\gamma(\eta=\eta_0,\theta=0))=-\delta\eta(\gamma(\eta=\eta_0,\theta=0))=-\delta\chi(\gamma(\eta=\eta_0,\theta=\pi))=  
\delta\eta(\gamma(\eta=\eta_0,\theta=\pi))$.
Furthermore, $\delta\theta$ and $\delta\phi$ for these light paths vanish because 
of axial-symmetry. These implies the intersection of 
$\gamma(\eta=\eta_{crit},\theta=0)$ and 
$\gamma(\eta=\eta_{crit},\theta=\pi)$  does not change its time $\eta$ 
but position $\chi$ by $2\delta\chi$ along the $z$-axis. 
Since there is no peak of $Y_{10}$ between $\theta=0,\pi$, also all the 
other $\gamma(\eta=\eta_{crit},\theta,\phi)$'s should be shifted so as to pass 
the same position of $2\delta\chi$ on the $z$-axis at $\eta=\eta_{crit}$. Therefore the original 
zero-dimensional SOEP is
only shifted in the $z$-direction by $2\delta\chi$. There is no change of 
the TOEH.

\begin{figure}
        \centerline{\epsfxsize=16cm \epsfbox{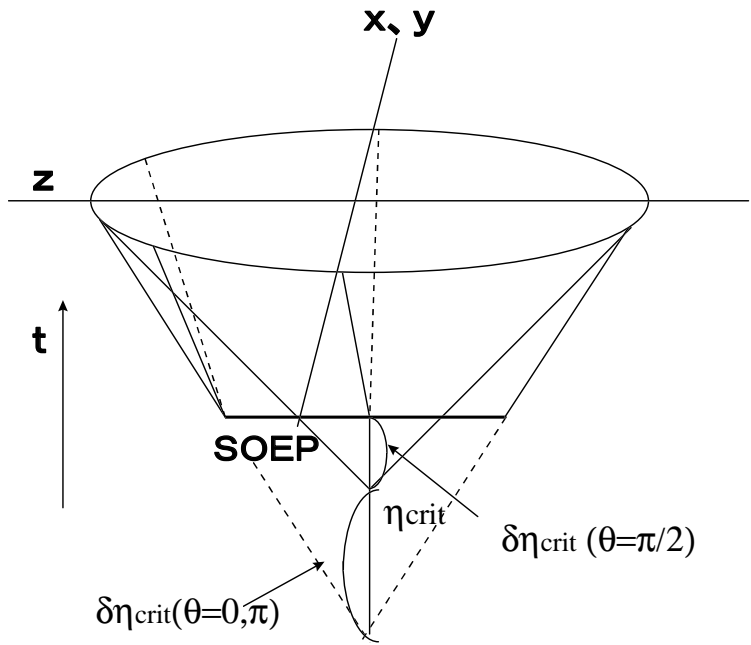}}
        \caption{The latest light paths with maximal $\delta\eta_{crit}$ ($x,y$
          direction in this figure) form an endpoint at the origin with
        $\eta=\eta_{crit}+\delta\eta_{crit}(\theta=\pi/2)$. On the other hand,
        a light path on the other axis ($z$-axis in this figure) crosses
        light paths from other directions and form an endpoint
        there. Thus the SOEP gets a dimension in this ($z$-) direction.}
        \protect\label{fig:pend}
\end{figure}

The even parity mode with $L=2$ possesses reflection symmetries about three
orthogonal planes. For small perturbation, these modes change the spherical 
wave front of light to the ellipsoidal one. By an appropriate rotation, the principal
axes of the ellipsoidal
wave front become $x$-, $y$-, and $z$-axis. Then, it is sufficient to determine
light paths along these axes. By the symmetry, $\delta\theta$ and
$\delta\phi$ vanish for these light paths. Since $\delta\eta=-\delta\chi$ 
means the change of $\eta_{crit}$ is given by 
$2\delta\eta(\gamma(\eta_{crit}))$, $\delta\eta_{crit}$'s of the light 
paths on the principal axes by each even 
parity $Y_{2M}$ mode are given by
\begin{eqnarray*}
  \delta\eta_{crit}(L=2,M=0)=\sqrt{\frac5\pi}H,
  -\frac12\sqrt{\frac5\pi}H,-\frac12\sqrt{\frac5\pi}H,\\
 \delta\eta_{crit}(L=2,M=1)=\frac32\sqrt{\frac5{6\pi}}H,
0, -\frac32\sqrt{\frac5{6\pi}}H,\\
 \delta\eta_{crit}(L=2,M=1)=\frac32\sqrt{\frac5{6\pi}}H,
0, -\frac32\sqrt{\frac5{6\pi}}H,
\end{eqnarray*}
where $H$ (the factor not depending on $Y_{LM}$) is given 
by eq. \ref{eqn:dec}.
By these results, we see the shape of the SOEP around the origin. Light 
paths from the latest direction (maximal $\delta\eta_{crit}$) form an endpoint 
at the origin (for example, see Fig.\ref{fig:pend}). On the other hand, light paths on the other axes will 
cross a light path from another direction not passing the origin, 
at a position different from the origin, so that their intersections provide the dimensions 
of the SOEP to their directions.
Thus, the case of $L=2, M=\pm 1,2$ provides two-dimensional SOEP. On the
contrary, the SOEP with $L=2, M=0$ depends on the signature of $H$. If 
$H$ is negative (positive), the SOEP is one (two)-dimensional (see Fig.\ref{fig:pend}). Since $H$ 
is generally not equal to zero, the TOEH is not stable under the 
perturbation with $L=2$.

By the mode with
$L>2$, the wave front will experiences more complicated deformation. By such a
deformation, the SOEP will get branching and become highly 
complicated as stated in the remark of the theorem\ref{T2}. For these 
modes,  $\delta\theta$ and $\delta\phi$ will not be
excluded from the discussion. A detailed investigation, however, would show the change of the
structure of the SOEP occurs even with non-vanishing $\delta\theta$,
$\delta\phi$.

\subsection{odd parity mode and higher order contributtion}
The metric perturbation of the odd parity mode is given by
\begin{equation}
        h_{ab}=\bordermatrix{
                   &
                   \eta,V \  \chi,U 
                   &
                   \theta \ \ \ \phi 
                    \cr
                    &\begin{array}{cc}
                        0 & 0  \\
                        0 & 0
                    \end{array}  &  \bar{h}_AS_\alpha  \cr
                    & Sym         &  \bar{h}S_{(\alpha;\beta)}\cr
        },
        \label{eqn:odd}
\end{equation}
where $S_\alpha$ is the transverse vector harmonics on the unit 
sphere\cite{SG}.
From (\ref{eqn:dec}) and (\ref{eqn:odd}), it is clear that the odd parity
mode does not affect $\delta\eta$, $\delta\chi$ in linear order. On
the other hand, though $\delta\theta$ and $\delta\phi$ exist, they do not
affect the structure of the SOEP. For, without $\delta\eta$ and
$\delta\chi$, all the perturbed outgoing light paths whose original past endpoint 
in background 
is the origin at $\eta=\eta_0$, start the origin at the same time $\eta_0$. They
still have only one endpoint at the origin with $\eta=\eta_{crit}$. 

For the modes not changing the structure of the SOEP, it would be 
necessary to investigate contributions from higher order evaluation. The higher order
contributions  are contained in the back-reaction of the changes of the
light path to the equation of null geodesics.
Nevertheless, it is also necessary to include second order
metric-perturbation. It will cause the difficulty of further
investigations.  In the second order, there should be mode coupling between
different parities, $L$'s, and $M$'s. This fact implies generally the
structure of the SOEP is unstable in the higher order. Even if so, however, there are the difference
of the sensitivity of the SOEP among each mode. The TOEH is insensitive to
odd parity mode and $L=1$ even parity mode.
\section{The structural stability of the topology of the event horizon}
In the previous section, it is shown that the spherical TOEH is unstable 
under the linear perturbation. Since there is no appropriate example of
a spacetime, however, with non-spherical topology, similar analysis is
impossible for other TOEHs.
Then, in this section we discuss the structural stability of the SOEP of 
the EH in catastrophe theory. As discussed in the section 2, it corresponds to the stability of the 
TOEH. First, we investigate it in a (2+1)-dimensional 
spacetime. 
\subsection{in (2+1)-dimensional spacetime}

The plan of analysis is following. First of all, we consider the appropriate wave
front of light in a flat spacetime. According to geometrical optics, 
the wave
front produces backward caustics and the endpoints of a null surface 
related to the wave front. In the context of catastrophe theory, Thom's
theorem state that the structures of such caustics are 
classified\cite{PS}, if they are 
structurally stable. So, we analyze the structure of the caustics and judge 
whether the SOEP is classified by Thom's theorem. Here, we consider that
the structural stability corresponds to the stability under the small change
of the shape of the wave front and the local geometry around the
endpoints. The shape of the wave front reflects the global structure of 
the spacetime between $\scri^+$ and the wave front. Furthermore, the stability is that of only the local structure 
of the caustics. Therefore, to discuss it in the flat spacetime is valid as 
long as we deal with the structure of a small neighborhood.

\begin{figure}
        \centerline{\epsfxsize=16cm \epsfbox{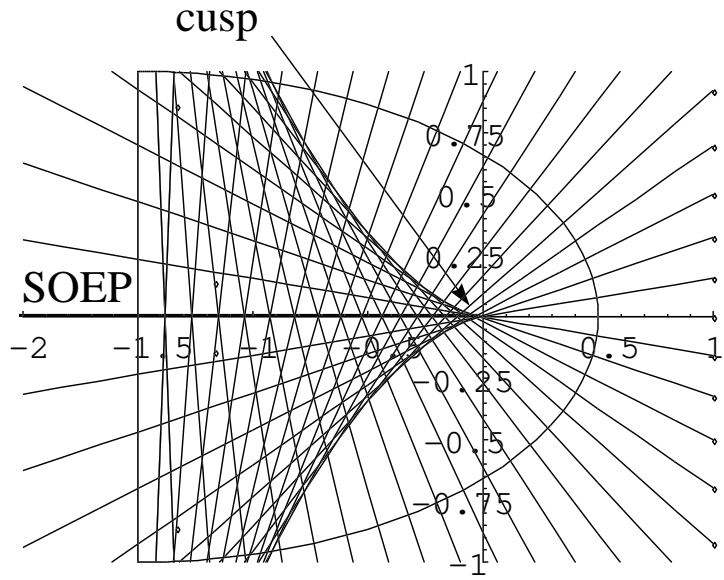}}
        \caption{The light paths for the elliptic wave frontwith $a=2,
          b=1$  are drawn. 
        There are the crossing points of the light paths which are the 
        endpoints of a null surface corresponding to the wave front, on 
        the $x$-axis, $[2x_0,0]$. A cusp is formed at the origin.}
        \protect\label{fig:cau2}
\end{figure}

For simplicity, we consider only the elliptical wave front, 
\begin{eqnarray}
               E_2:\  \left({x-x_0\over a}\right)^2+\left({y\over b}\right)^2=1\\
                x_0=-{a^2-b^2\over a},\ \ \  a\geq b\ .
        \label{¥}
\end{eqnarray}
Then, the square of the distance between ($x,y$) and ($X,Y$) is given by
\begin{eqnarray}
        f_{XY}\left({\bf x}\right)&=&\left(X-x\right)^2+\left(Y-y\right)^2\\
               &=&\left(X-\left(a \sqrt{1-\left(y/b\right)^2}+{b^2-a^2\over a}\right)\right)^2+\left(Y-y\right)^2,
        \label{}
\end{eqnarray}
where ($X,Y$) is an arbitrary point and $x-x_0$ is positive.
As known in geometrical optics, in a flat spacetime a light path through
($X,Y$) is 
given by the stationary points of $f_{XY}\left(x,y\right)$;
\begin{eqnarray}
        {\partial f_{XY}\left({\bf x}\right)\over \partial y} & = & 0 \ \ \ \Rightarrow
        Y  =  A(y) X +B(y)
        \label{¥} \\
        A(y)&=&-\left({\partial x(y)\over \partial y}\right)_{E_2} , \ \ 
        B(y) = x\left({\partial x(y)\over \partial y}\right)_{E_2}+y,
        \label{¥}
\end{eqnarray}
where $(\partial/\partial)_{E_2}$ means partial derivative with a constraint $E_2$.
The light paths are drawn in Fig.\ref{fig:cau2}. From this figure, we see 
that
they form a caustic at the origin, and the SOEP of the null surface 
concerning the wave front is a one-dimensional set, an interval on the 
$x$-axis  $[2x_0,0]$.

To see the structure of the caustic, we derive the Taylor 
series of $f_{XY}\left({\bf x}\right)$ around the origin,
\begin{eqnarray}
        \widetilde{f_{XY}}\left({\bf x}\right) & = & \left({b^4 \over a^2}-{2 b^2 X \over a}+X^2+Y^2\right)-
        2 Y y +{a X y^2 \over b^2} \\
         & + & \left({a^2 \over 4 b^4}-{1 \over 4 b^2}+{a X \over 4 
         b^4}\right)y^4+O\left(y^5\right).
        \label{}
\end{eqnarray}
Hence, the light paths form a cusp (a type $A_3$ catastrophe) at $(X,Y)=(0,0)$ because of 
$\widetilde{f}\sim y^4$. 
From Thom's theorem, it is structurally stable except for $a=b$. Of course, $a=b$ corresponds the
circular wave front and the zero-dimensional SOEP.

\subsection{in (3+1)-dimensional spacetime}
\begin{figure}
        \centerline{\epsfxsize=16cm \epsfbox{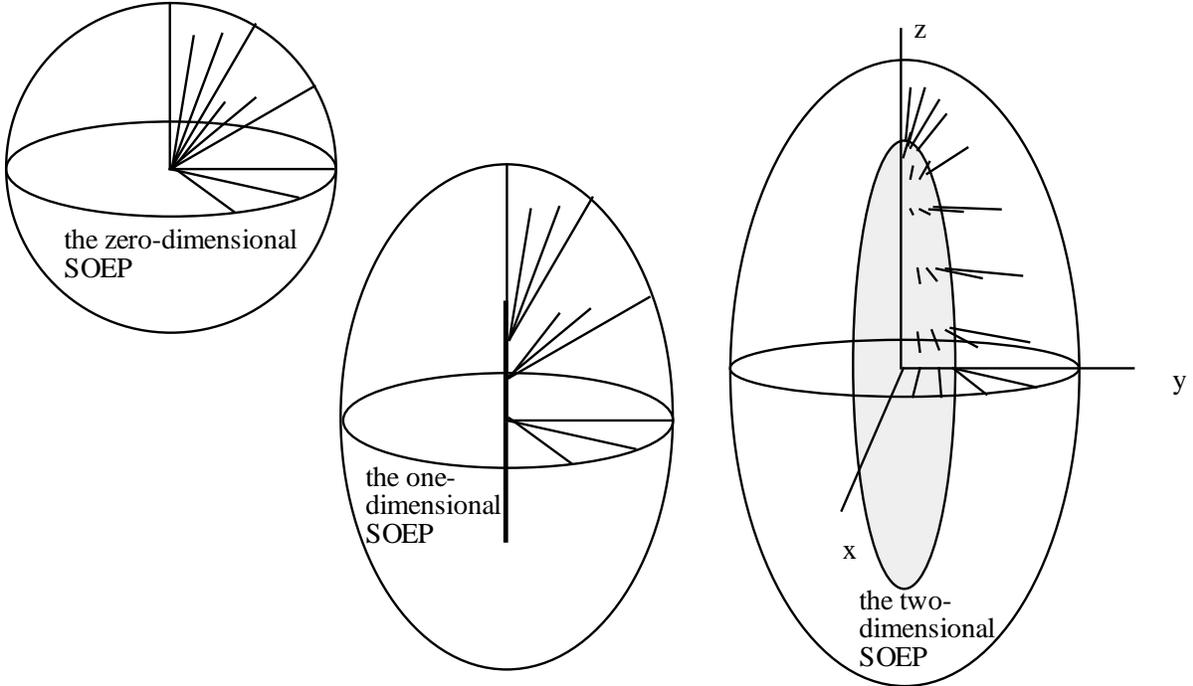}}
        \caption{The SOEP becomes zero-dimensional for the spherical wave 
        front. In the prolate-spheroidal wave front, the one-dimensional 
        SOEP appears. Otherwise, the ellipsoidal wave front produces the 
        two-dimensional SOEP.}
        \protect\label{fig:eli}
\end{figure}

\begin{figure}
        \centerline{\epsfxsize=16cm \epsfbox{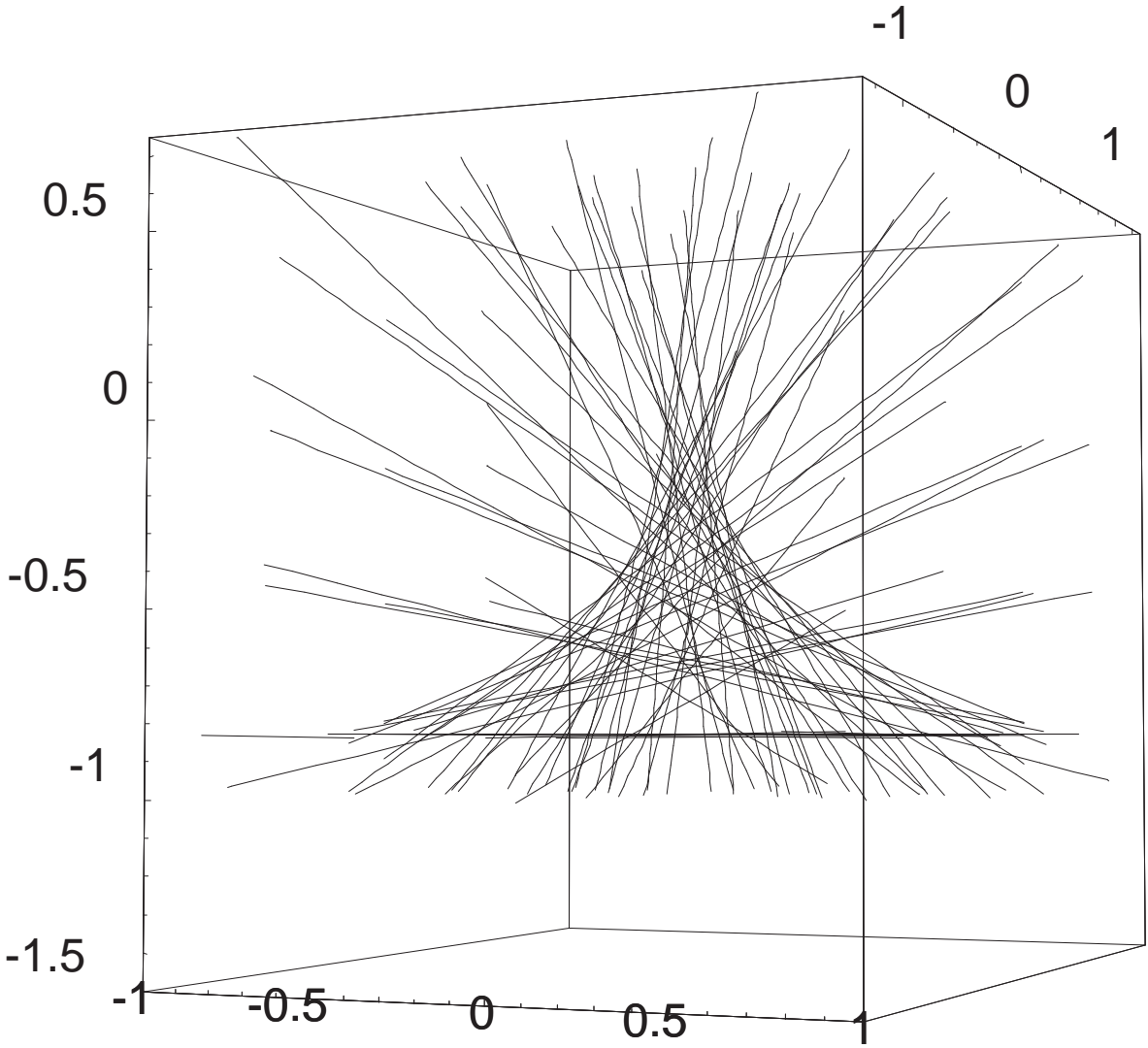}}
        \caption{The light paths for the ellipsoidal wave front with
          $a=1, b=1.3, c=1.5$ are 
        drawn.  A cusp is formed at the origin. Watching this figure 
        carefully, one will see that also another cusp exists at $(0,0,(b^2-a^2)/c)$}
        \protect\label{fig:cau3}
\end{figure}

For a (3+1)-spacetime, the investigation above can similarly
be done, though its situation becomes a little complex. In this case, 
there are 
three possibilities of the SOEP of the EH, even after sufficient 
simplification. As shown in Fig.\ref{fig:eli}, the endpoint forms a 
point, line, or surface. As the previous subsection, we consider the ellipsoidal
wave front,
\begin{eqnarray}
              E_3:  \left({x\over a}\right)^2+\left({y\over b}\right)^2+\left({z-z_0\over c}\right)^2=1\\
                z_0=-{c^2-a^2\over c},\ \ \  0< a\leq b\leq c.
        \label{}
\end{eqnarray}
For a $z-z_0>0$ branch, the square of the distance between $(x,y,z)$ and 
an arbitrary point $(X,Y,Z)$ is given by
\begin{eqnarray}
        f_{XYZ}\left({\bf x}\right)&=&\left(X-x\right)^2+
        \left(Y-y\right)^2+\left(Z-z\right)^2\\
               &=&\left(X-x\right)^2+\left(Y-y\right)^2+
               \left(Z-(c\sqrt{1-\left(x/a\right)^2-\left(y/b\right)^2}+
               z_0)\right)^2.
        \label{}
\end{eqnarray}
The light path through $(X,Y,Z)$ is given by
\begin{eqnarray}
        {\partial f_{XYZ}\left({\bf x}\right)\over \partial x} & = & 0,\ \ 
        {\partial f_{XYZ}\left({\bf x}\right)\over \partial y} = 0 
        \label{} \\
        \Rightarrow X & = & A(x,y) Z +B(x,y),\ \ \  and\ \   Y=C(x,y) 
        Z+D(x,y),
        \label{} \\
        A&=&-\left({\partial z(x,y)\over\partial x}\right)_{E_3} , \ \ 
        B =z\left({\partial z(x,y)\over\partial x}\right)_{E_3}+x,\ \  \\
        C&=&-\left({\partial z(x,y)\over\partial y}\right)_{E_3},\ \  
        D=z\left({\partial z(x,y)\over\partial y}\right)_{E_3}+y.
        \label{}
\end{eqnarray}
From Fig.\ref{fig:cau3} showing the light paths, it is known that a caustic  
is formed around
 the origin. Only when $a$, $b$ and $c$ are equal to each other, the SOEP
 becomes
zero-dimensional (at the origin). $a=b\neq c$ implies the endpoints form
a one-dimensional set which is an interval on the $z$-axis,
 $[2z_0,0]$. Otherwise, the SOEP is two-dimensional (Fig.\ref{fig:eli}).

The Taylor series of the potential $f$ at the origin is given by
\begin{eqnarray}
        \widetilde{f_{\bf X=0}}\left({\bf x}\right) &=& {a^4\over c^2}+
        {-a^2+c^2\over 4 a^4}x^4+
{-a^2+c^2\over 8 a^6}x^6
          +  {b^2-a^2\over b^2}y^2+{-a^2+c^2\over 2 a^2 b^2}x^2 y^2\\ &+&
         {3\left(-a^2+c^2\right)\over 8a^4 b^2}x^4y^2 + {-a^2+c^2\over 
         4 b^4}y^4+{3\left(-a^2+c^2\right)\over 8a^2b^4}x^2y^4+{-a^2+c^2\over 
         8b^6}y^6\\ &+& O\left({\bf x}^7\right).
        \label{}
\end{eqnarray}
The structure of the caustic is controlled by the leading term of
$\widetilde{f}$ about $x,y$. 
When $a<b\le c$, $\widetilde{f}\sim\alpha x^4 + \beta y^2+\gamma$ produces a cusp (type $A_3$).
Then, the two-dimensional SOEP is structurally stable. On the other hand,
 $\widetilde{f}$ becomes 
$\alpha (y^4+2x^2y^2+x^4)+\gamma$ with $a=b\neq c$. This case corresponds to 
the line SOEP and it is 
not structurally stable. Incidentally, if $a<b<c$, there is also another 
cusp at $(0,0,-(b^2-a^2)/c)$ (carefully see Fig.\ref{fig:cau3}). The 
Taylor expansion of $f$ around this cusp tells that it is also stable as 
long as $a\neq b$ and $b\neq c$. With $b\rightarrow a$, this cusp 
approaches the cusp at the origin and degenerate into the unstable 
structure. On the contrary, when $b$ is equal to $c$, this cusp 
disappears at the center of the ellipsoid. The example given in \cite{ST} 
corresponds to this case. Of course, the zero-dimensional SOEP $(a=b=c)$  
is not structurally stable.

\section{Summary and Discussions}
We have investigated the stability of the topology of the EH (TOEH).
First the stability of a spherical topology is
investigated under linear perturbation in a spherically symmetric
background. In linear order, $L=2$ even parity mode changes the
structure of the SOEP and the TOEH, and odd parity mode and $L=1$ mode
do not. In higher order, however, mode coupling between the  modes with 
different $L$ or parity will cause the
instability of the TOEH even for these modes not changing the TOEH in the 
linear order. For $L> 2$ even parity mode, more detailed
investigation will be required. Anyway, we have seen that the trivial TOEH
is generally unstable under the linear perturbation. 
 
In this discussion of the linear perturbation, we consider
Oppenheimer-Snyder spacetime as the example of an always spherical
EH. Nevertheless, the result will be same to other non-eternal EHs with
spherical symmetry since we have never used the concrete geometrry of
the spacetime other than the spherical symmetry.

How can we interpret the fact that the TOEH is insensitive to some
modes of the perturbation? In a sense, when we give  odd parity or $L=1$
 perturbation to the spherical EH, the
change of the TOEH in the higher order would not be able to be detected
(though it is not trivial how one can observe it). For, while a local geometry around an observer is perturbed with
the same strength as the given perturbation,
the TOEH is not so. The change of the TOEH in the higher order would be prevented being observed.

Second, by
a simple discussion in catastrophe theory, the structural stability of 
the SOEP is 
studied in a more general situation. Assuming the ellipsoidal wave front, 
the stability of zero-, one-, and two-dimensional SOEP is
investigated. We see that the two-dimensional SOEP is stable, and the
one- and zero-dimensional is not. Therefore, the TOEH with handles
(a torus, a double torus, ...) is generic.

Though in the present article we meet only the SOEP with a cusp 
catastrophe, as discussed
 in \cite{AR} there 
will be the possibilities of some further types, the `swallowtail', the
`pyramid' , and so on. They will form other structurally stable
SOEPs. The TOEH with them will be revealed in our forthcoming work.

One may expect to give some restriction to the TOEH 
introducing certain conditions about matter field.
The present results imply, however, that it is likely hopeless. It seems that the symmetry of the spacetime affect
the structure of the SOEP of the EH and the TOEH, and is
easily disturbed by perturbation.
If one does not concern the scale of the topological structure of the EH, the TOEH 
can generally become complicated.

\centerline{\bf Acknowledgements}

We would like to thank Dr. K. Nakao, Dr. H. Susa and Professor Y. Jun-ichi 
for helpful discussions. We are grateful to Professor H. Sato and 
Professor N. Sugiyama for their continuous encouragement.
The author thanks the Japan Society for the 
Promotion of Science for financial support. This work was supported in 
part 
by the Japanese Grant-in-Aid for Scientific Research Fund of the Ministry 
of 
Education, Science, Culture and Sports.

\end{document}